\documentclass [prl, twocolumn, showpacs,a4paper]{revtex4}
\usepackage[pdftitle={Coherent vibrations of submicron spherical gold shells in a photonic crystal},
colorlinks=true,
breaklinks=true,citecolor=blue,pdfkeywords={ultrafast, coherent
acoustic waves, core-shell particles, plasmon, photonic crystal,
Lamb mode},a4paper=true]{hyperref}

\usepackage{graphicx}
\usepackage{bm}
\usepackage{dcolumn}
\usepackage{natbib}

\begin{document}

\title{Coherent vibrations of submicron spherical gold shells in a photonic crystal}

\author{D.~A.~Mazurenko}
\altaffiliation[Present address: ]{Zernike Institute for Advanced
Materials, University of Groningen, Nijenborgh 4, 9747 AG Groningen,
the Netherlands} \email{D.A.Mazurenko@rug.nl}
\author{X.~Shan}
\author{J.~C.~P.~Stiefelhagen}
\author{C.~M.~Graf}
\altaffiliation[Present address: ]{ Institut f\"{u}r Chemie und
Biochemie, Freie Universit\"{a}t Berlin, Takustr. 3, 14195 Berlin,
Germany }
\author{A.~van~Blaaderen}
\author{J.~I.~Dijkhuis}
\email{J.I.Dijkhuis@phys.uu.nl}

\affiliation{Debye Institute, Department of Physics and Astronomy,
University of Utrecht, P.O. box 80000, 3508 TA Utrecht, the
Netherlands}

\date{\today}

\begin{abstract}
Coherent acoustic radial oscillations  of  thin spherical gold
shells of  submicron diameter excited by an ultrashort optical pulse
are observed in the form of pronounced modulations of the transient
reflectivity on a subnanosecond time scale. Strong acousto-optical
coupling in a photonic crystal enhances the modulation of the
transient reflectivity up to 4\%. The frequency of these
oscillations is demonstrated to be in good agreement with Lamb
theory of free gold shells.
\end{abstract}


\maketitle Acoustic motion in nanoscale objects driven by light has
attracted considerable attention over the last decade.  The interest
is explained by various potential applications in nanomechanics,
like nanomotors \cite{Eelkema06}, ultrahigh-frequency acoustic
oscillators \cite{Huang03}, and acousto-optic modulators
\cite{Okawachi05}.

Vibrational modes confined in nanoparticles can be excited by a
short optical pulse and observed as  modulations of the transient
reflectivity or transmission on a picosecond timescale
\cite{Hartland06}. Up to now  such vibrations have only been
observed in  the core of solid particles. Recent progress in
fabrication of monodisperse multicoated metallo-dielectric colloids
\cite{Averitt97,Oldenburg98,Graf02}, arranged in a periodic fashion
and forming a photonic crystal \cite{Graf03} makes vibrations
localized in thin shells accessible for experiments.
 Excitation of acoustic vibrations in
such  structures has   two important aspects: First, a shell
requires much less vibrational energy than a massive sphere would to
reach equal optical responses. Second, the  photonic order may
enhance the acousto-optic coupling.  Thermal quadrupolar
hollow-shell vibrations of nickel-silver core-shell nanoparticles
have been recently observed in Raman scattering experiments
\cite{Portales02}. However, to the best of our knowledge optical
excitation of ground-state oscillations localized in a shell has
never been shown.
 In
this Rapid Communication we demonstrate optical excitation of
\textit{coherent radial oscillations} of thin gold shells covering
inner silica cores of submicrometer diameter. We determine the
intrinsic lifetime of these Lamb modes.

\begin{figure}[b] \centering
\scalebox{1.00}{\includegraphics{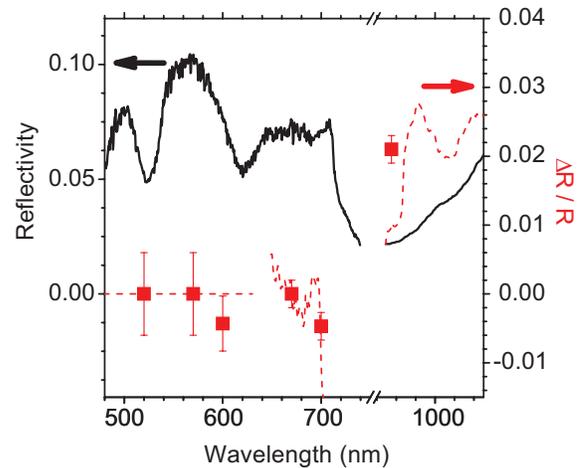} }
\caption{\label{linspec} (Color online) Measured linear reflection
spectrum of an ordered 3D array of silica-core gold-nanoshells with
a silica outer shell (black solid line). Gray (red) squares show the
measured amplitude of oscillations in the transient reflectivity,
$A_1$, dashed line the calculated  values $\delta R /R$: according
to Eq.~\ref{eq4} for $\lambda >650$~nm and $\delta R /R = 0$ for
$\lambda < 600$~nm, respectively. }
\end{figure}

Our particles consist of a $228$-nm radius silica core, a gold shell
with a thickness of $38$~nm, and an outer silica shell with a
thickness of $10$~nm. The particle size polydispersity is $<5\%$ as
deduced from transmission electron microscopy (TEM) pictures
\cite{Graf03}. The particles were assembled in close-packed ordered
three-dimensional (3D) arrays thus forming a metallo-dielectric
photonic crystal that serves to enhance acousto-optical coupling.
The details of the particle synthesis can be found in
Ref.~\cite{Graf03}. For our studies we select a
 highly ordered region on the sample, where the photonic crystal is thicker
than the penetration depth of the light and the reflection from the
substrate is negligible. Our structure possesses  spatial
periodicity for both acoustic and optical properties. Since the
spheres are in mechanical contact only in few points, the acoustic
interaction of adjoining spheres is expected to be small and further
neglected. However, the periodic arrangement is important for the
electromagnetic waves and here serves to enhance the acousto-optical
coupling. The black solid line in Fig.~\ref{linspec} shows a typical
linear reflection spectrum of a highly ordered part of our photonic
crystal. The spectrum has
 several resonances, which appear to be  much sharper than for a
 dilute array of the same but in this case optically uncoupled
 particles \cite{Graf02}. In our sample we distinguish two kinds of
 resonances
\cite{MySPIEAu, Mythesis, Kelf05}: First, the so-called Bragg
resonance for wavelengths  close to the lattice spacing parameter,
$\lambda \sim 576$~nm. The spectral position of this resonance is
strongly dependent on the incident angle of the incoming light.
Second, collective plasmonic Mie resonances, which are independent
of the angle of insidence and dominate the reflection spectrum  for
$\lambda > 650$~nm. These collective Mie resonances, however, are
coupled with Bragg modes, particularly for $\lambda <650$ nm.

Our sample was  excited by a $120$-fs pulse extracted from an
$800$-nm     amplified Ti-sapphire laser operating at $1$~kHz. The
pump pulse was focused onto a \mbox{400-$\mu $m} spot at the sample
surface with an energy density of $\sim 0.5$~mJ/cm$^{2}$ per pulse.
This density is close to but slightly below the damage threshold and
at least one order of magnitude less than the excitation power used
to reach the melting temperature   in solid spheres \cite{Plech04}.
The transient reflectivity was probed by a white light continuum
generated by a beam split-off from the same laser and focused either
in a cuvette filled with acetone or in a sapphire plate. The
white-light pulse was passed via an optical delay line and focused
onto a \mbox{25-$\mu $m} spot at the sample surface within the
illuminated area of the pump. The reflected light was subsequently
dispersed in a spectrometer and registered by a charge-coupled
device (CCD). Temporal evolutions of the reflectivity integrated
over  a selected  30-nm bandwidth and as a function of delay were
detected by an  InGaAs photodetector equipped with an amplifier with
0.1~ms response. The signal from the photodetector was integrated by
a digital voltmeter over 10~$\mu $s. All experiments have been
carried out at room temperature.

\begin{figure}[t] \centering
\scalebox{0.95}{\includegraphics{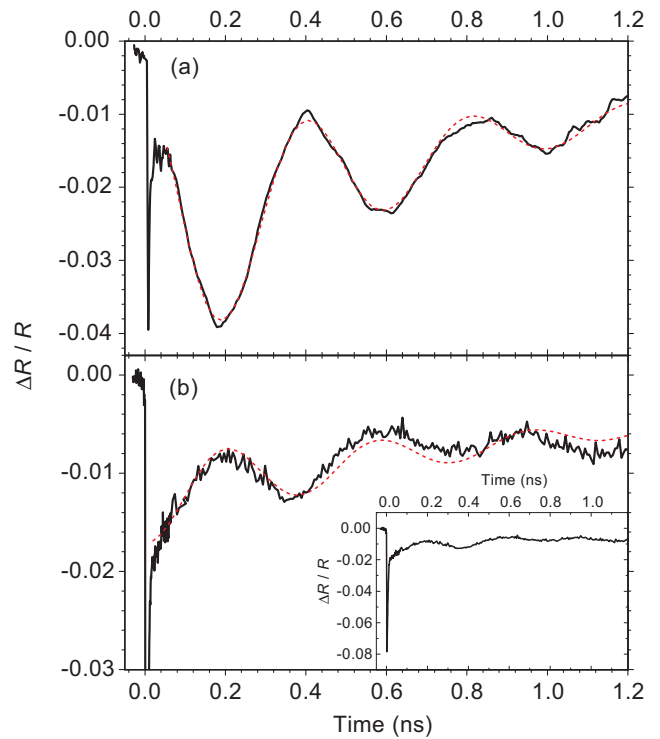} }
\caption{\label{kinet} (Color online) Transient reflectivity
dynamics of the gold-shell array measured at (a) $950$~nm and (b)
$700$~nm. Black solid lines depict experimental data, red dotted
lines show the fits. Inset shows the signal measured at $700$~nm
plotted including the entire electronic contribution.}
\end{figure}

 The dynamics of the transient reflectivity is found to be
 dependent on the selected probe wavelength.
 In Fig.~\ref{kinet}, the black solid lines
show evolutions of the transient reflectivity, $\Delta R / R$,
registered
 at (a) $950$~nm   and (b) $700$~nm.
The inset in Fig.~\ref{kinet}(b) shows the signal at $700$~nm at
full scale. Unfortunately, the spectral range of $740-950$~nm was
not accessible for measurements because of intense scattering of the
 $800$-nm pump beam. Both curves have a large and sharp peak of
picosecond duration  immediately after the pump.
 On a subnanosecond timescale the
transient reflectivity shows a quite distinct behavior. In both
Fig.~\ref{kinet}(a) and (b), we observe pronounced oscillations of
the reflectivity with a period of about $400$~ps,  independent of
the probe wavelength, $\lambda$. The amplitudes of these
oscillations, however, are dependent on $\lambda$ and reach an
amplitude   as much as $4\%$ of the total reflected intensity at
$\lambda = 950$~nm [Fig.~\ref{kinet}(a)]. In a disordered sample of
the same batch of particles, however, we were not able to measure
any oscillations. We explain this phenomenon by the fact that
optical resonances in our photonic crystal are much sharper than in
 arrays of individual gold-shell spheres
\cite{Averitt97,Oldenburg98,Graf02} and, as a result, exhibit a much
stronger acousto-optical coupling.  At $\lambda = 700$~nm, the
amplitude of the oscillations is smaller but still quite sizable.
Further, we did observe  weak oscillations at $600$~nm (not shown in
Fig.~\ref{kinet}) but we found no oscillations at other wavelengths.
It is interesting to note that at $950$~nm and $700$~nm the initial
peaks have the same signs,  while the slow oscillations have
opposite polarities. This directly shows that the fast spike and the
slow oscillations must have different origins. The temporal
evolution of the signal can be approximated quite faithfully by the
function
\begin{equation}
\label{eq1} \frac{\Delta R}{R} = -A_1 \exp \left( {-t / \tau _1 }
\right)\cos \left( {\frac{2\pi }{T}t - \varphi } \right) + A_2 \exp
\left( {-t / \tau _2 } \right).
\end{equation}
\noindent Here, $t$ is time and the fitting parameters $T$ and
$\varphi$ are the period and the phase of the oscillations,
respectively, and $\tau _1$ and $\tau _2$  decay times. Further,
$A_1$ and $A_2$ are amplitudes referring to the oscillatory and
non-oscillatory decay, respectively.
 The best fits of  $\Delta R / R$ for $950$~nm and $700$~nm  are
shown by dotted lines in Fig.~\ref{kinet}(a) and (b), respectively.
The  results are collected in Table~\ref{tabfit}. Clearly the period
and the phase of the oscillations are virtually constant over the
full spectral range as are  $\tau _1$ and $\tau _2$. The lowest line
of Table~\ref{tabfit} collects the average values of all fitting
parameters over different wavelengths and points on the sample. We
conclude that the detected phase of the oscillations is zero radians
and the period of oscillations is $390$~ps with a standard deviation
of $5\%$.

\begin{table*}
\caption{\label{tabfit}  Fitting parameters of the transient
reflectivity using Eq.~(\ref{eq1}) for different wavelengths. Last
row summarizes the average values from different measurements.}
\begin{ruledtabular}
\begin{tabular}{cdddddd}
\multicolumn{1}{c}{$\lambda $ (nm)}&\multicolumn{1}{c}{$ A_1
$}&\multicolumn{1}{c}{$ A_2 $}
&\multicolumn{1}{c}{$\tau _1$ (ps)}&\multicolumn{1}{c}{$\tau _2$ (ps)}& \multicolumn{1}{c}{$T$ (ps)}& \multicolumn{1}{c}{$\varphi $ (rad)}\\
\hline
950&-0.021&-0.028&482&1176&406&0.0 \\
700& 0.0047&-0.015&633&1510&378&0.08\pi  \\
600& 0.0043&-0.025&770&> 1000&381&-0.06\pi \\
\multicolumn{1}{c}{Average}& \multicolumn{1}{c}{depends on
$\lambda $} & \multicolumn{1}{c}{depends on $\lambda $} &
\multicolumn{1}{c}{$600 \pm
200$} &\multicolumn{1}{c}{$1300 \pm 300$} &\multicolumn{1}{c}{$390 \pm 20$} &\multicolumn{1}{c}{$0 \pm 0.1\pi$}\\

\end{tabular}
\end{ruledtabular}
\end{table*}

 The initial peak in the transient reflectivity is caused by hot
electrons in gold. The subsequent dynamics is due to equilibration
of the electron gas with the lattice~\cite{Hartland06} and takes no
longer than $20$~ps. The nature of the oscillations can not be found
in electron-temperature variations. We attribute the $390$-ps
oscillations  in the transient reflectivity to induced
\emph{coherent acoustic vibrations of the submicron gold-shells}
following a  rapid change in lattice temperature of the gold shell.

\begin{figure}[t] \centering
\scalebox{0.95}{\includegraphics{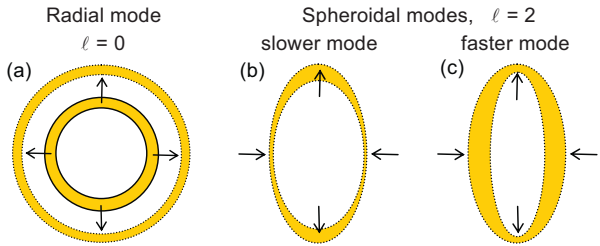} }
\caption{\label{fig:modes}  Sketch of  (a) $l=0$ (radial), and (b)
and (c) two types of $l=2$ spheroidal vibration modes of a hollow
shell sphere.}
\end{figure}

 In order to determine the eigenfrequencies of the acoustic
vibrations of the particle we assume that the acoustic coupling of
the gold shell to the silica core and outer shell is \emph{weak} and
the acoustic response in our particles can be modeled as that of a
free-standing thin hollow sphere. This approach is justified by the
substantial acoustic mismatch between silica and gold and the weak
mechanic contact between core and shell. Indeed, the thermal
expansion coefficient is much higher for gold than for silica and,
therefore, at elevated temperature the gold shell is not in contact
with the silica core. Vibrational modes of a thin shell are
classified into two categories - torsional and spheroidal modes
  \cite{Lamb82sh}, of which only the even-$l$ spheroidal modes are
optically active \cite{Duval92}. Of all even-$l$ modes, the most
important ones are expected to be the $l=0$ and the $l=2$ spheroidal
modes because they possess the highest symmetry and therefore
optical coupling. A sketch of these modes is presented in
Fig.~\ref{fig:modes}.

 Assuming zero
tension on the interfaces, the period of the Lamb oscillations can
be expressed \cite{Lamb82sh} for the ground  $l=0$ mode as
\begin{equation}
\label{eq2} T_0 =  \pi  \xi^{-1/2} r_s / c_t
\end{equation}
\noindent and for $l=2$ mode as
\begin{equation} \label{eq3}
T_{2\pm } = 2\pi  \left[ {5 \xi + 2 \pm \left( {25\xi^2 +4\xi+4 }
\right)^{1/2}} \right]^{-1/2} r_s/c_t.
\end{equation}
\noindent Here,  $\xi = 3-4\left(c_t/c_l\right)^2$, $c_t=1200$~m/s
and $c_l=3240$~m/s are the longitudinal and transverse sound
velocities of gold, respectively. Further,  $r_s=247$~nm is the
average radius of the gold shell.
 Using Eqs.~(\ref{eq2}) and (\ref{eq3}) we obtain $T_0=413$~ps for the $l=0$
 mode.  The exact solution for $38$~nm thick shell
 gives $T_0=411$~ps confirming the validity of the thin shell
 approximation.
 For lower- and higher-frequency branches of the $l=2$ we find $T_{2+}=1074$~ps and $T_{2-}=249$~ps, respectively.
 Taking into account a
 5-\% spread in the measured period at different locations on the
 sample, we arrive at the conclusion that the calculated $l=0$  mode
 ($T_0=411$~ps) is within the experimental error from
 the value found in the experiment, $T=390$~ps, while the slower $l=2$ mode is  too slow.
  We also checked that oscillations of the silica core are too  fast
to explain the
 experiment: Using equations for the acoustic vibrations of a solid sphere \cite{Lamb82sp} with $c_t=3760$~m/s
and $c_l=5970$~m/s for silica,
 we  found a period of $145$~ps for the lowest  Lamb mode.

Oscillation of the gold shells are the result of the rapid increase
of the gold lattice temperature $\Delta \mathcal{T}$ and associated
thermal stress induced by hot electron-phonon relaxation. This is
consistent with the observed zero
 phase of the optical oscillations (see Table~\ref{tabfit}).

 Portales \emph{et al.}  studied  resonant Raman scattering from
 nickel-silver core-shell particles and found that their  spectra can be explained quantitatively
   by   thermal $l=2$ vibrations
 of the silver shell, i.e. assuming a stress-free  internal boundary conditions at the
 core interface \cite{Portales02}. This mode, however, is not observed in our
 pump-probe experiment. For solid spheres the difference between Raman and
 pump-probe experiments is well-known
 \cite{Courty02,Portales02,Bachelier04}. In Raman
scattering measurements  excitation is thermal and  modes are
occupied according to a Planckian  distribution. Since Raman
scattering is primarily sensitive to dipolar plasmon coupling with
the  modulation of the surface charges induced by a quadrupole
vibration ($l=2$) of the sphere, the $l = 2$ peak prevails. In
contrast,  pump-probe experiments are impulsive and after a time
much shorter than the period of the acoustic oscillations, electrons
are expected to reach a thermal equilibrium distribution in the
entire volume promoting the excitation of $l=0$ mode. Indeed, the
penetration depth of hot electrons in gold is $\sim 300$~nm, which
is comparable to a quarter of the gold-shell circumference
\cite{Maznev97}. As a result, the $l=0$ mode is predominantly
excited also in our relatively large particles.

 The oscillation amplitude of  $\Delta R / R$ can be
estimated by the following simple model.  As  already mentioned, the
spectral peak near 576~nm is due to  Bragg scattering in the
photonic lattice. Since the acoustic vibrations of the gold sphere
do not affect the lattice parameter of the gold-shell photonic
crystal we expect that the oscillation of $\Delta R / R$  vanish
near the Bragg resonance (left red dashed line in
Fig.~\ref{linspec}). In the red and infrared spectral ranges
($\lambda
>650$~nm), however, the spectral features are due to plasmon
resonances.
Periodic contractions and dilations of the gold shell lead to a
periodic modulation of the dielectric constant of gold,
$\varepsilon $, and thus shifts  the plasmon resonances back and
forth. The modulation of transient reflectivity caused by the
acoustic oscillations of shell can be expressed as
\begin{equation}
\label{eq4} \frac{\delta R}{R}=\frac{\partial R}{R\partial
\lambda}\left(\frac{\partial \varepsilon}{\partial
\lambda}\right)^{-1} \delta \varepsilon,
\end{equation}
where $\delta$ denotes the  oscillation amplitude. The dielectric
constant of gold can be expressed as a sum of the interband and
intraband Drude terms, $ \varepsilon =\varepsilon ^{i}-\omega
^{2}_{p}/ \omega ^{2}$, with $\omega = 2\pi c/\lambda$ optical
frequency, $c$ speed of light in vacuum, $\omega _{p}$ the plasma
frequency, which is proportional to the square root of the electron
density, $\omega _{p} \propto \sqrt{n}$. For sufficiently long
wavelengths ($\lambda > 650$~nm), we neglect $|\delta \varepsilon
^{i}| \ll |\delta \varepsilon| $\cite{Garfinkel66},  and the
modulation of the dielectric constant reads $\delta \varepsilon = -
\left(\omega _p/\omega\right)^{2}\left(\delta n / n\right)$. Noting
that $n$ is inversely proportional to the volume we obtain
 $\delta n /n = - 3\alpha \Delta \mathcal{T}$, with
$\alpha = 1.42\times 10^{-5}$~K$^{-1}$ the linear thermal expansion
coefficient \cite{Nix41}. The rise in the gold  temperature, $\Delta
\mathcal{T}$, in turn, can be estimated from  the electron-phonon
equilibration dynamics in a framework of a two-temperature model.
Analysis of  $\Delta R/R$ kinetics on the picosecond timescale in
terms of  rapid cooling of the electron gas and heating of the gold
lattice \cite{Hartland06} allows us to estimate $\Delta
\mathcal{T}=100\pm 50$~K. In Fig.~\ref{linspec} the middle and right
red dashed lines show the estimated $\delta R /R$, which appear to
be in a qualitative agreement with the experimental data (red solid
squares).

The observed decay of the  oscillations of $\Delta R/R$ cannot be
explained by dephasing caused by inhomogeneous variations in the
thickness or diameter of the gold shells. Indeed, the oscillation
period of a thin gold shell is independent of the shell thickness.
Further, if we assume that the particles are normally distributed
with a standard deviation $\sigma _r \ll r_s$, then for $t \ll Tr_s
/ \sigma _r$ the inhomogeneous decay of the oscillation amplitude of
$\Delta R/ R$ can be expressed as \cite{Hartland02}
\begin{equation}
\label{eq22}   S\left(t \right) \propto \cos \left(  2 \pi t/T
\right)\exp \left[ -\left( t / \tau _d\right)^2 \right],
\end{equation}
 \noindent with $\tau _d=r_s T/\sqrt{2}\pi \sigma _r$. Inserting $T
= 390$~ps and $\sigma _r /  r_s = 0.05$  known from the TEM data
\cite{Graf03} we obtain  $\tau _{d} = 1.75$~ns, which is three times
longer than the experimentally observed decay $ 0.6\pm 0.2$~ns.
Therefore, this inhomogeneous dephasing  mechanism is too slow to
fully explain the data. We believe that the decay of oscillations
can be explained by residual coupling of the radial $l=0$ mode with
other acoustic modes.

In conclusion,   the room-temperature transient reflectivity of a
photoexcited  silica-gold multishell photonic crystal exhibits
pronounced oscillations up to the nanosecond time scale. High
acousto-optical coupling in our photonic crystal in the red and
infrared serves to reach oscillation peak-to-peak amplitude  as high
as $4\%$ of the total reflectivity at moderate pump power. These
oscillations are caused by coherent,  radial vibrations of the
gold-shells. The frequency of the acoustic vibrations is found in
good agreement with  classical Lamb theory assuming free boundary
conditions on both sides of the shell. The damping of the ground
Lamb mode was shown to occur on a subnanosecond time scale and
points to a weak interaction with other acoustic modes. Propagation
of acoustic waves in a periodic array is an interesting point for
future experiments. Of particular interest is the  band of acoustic
modes between the lowest $l=2$ and the fundamental $l=0$ modes,
which is specific for spherical shells. Our result can be useful for
various acousto-optical applications, like fabrication of
high-frequency band-pass acoustic filters and switching of light
propagation in photonic crystals by acoustic waves.

We acknowledge  D.B.~Murray for his valuable comments on a draft of
this paper. We are grateful to C.R.~de~Kok, P.~Jurrius, and
P.~Vergeer for their technical assistance and A. Meijerink for
loaning us the CCD.

\bibliographystyle{apsrev}
\bibliography{Lamb_bib}

\begin{thebibliography}{22}
\expandafter\ifx\csname natexlab\endcsname\relax\def\natexlab#1{#1}\fi
\expandafter\ifx\csname bibnamefont\endcsname\relax
  \def\bibnamefont#1{#1}\fi
\expandafter\ifx\csname bibfnamefont\endcsname\relax
  \def\bibfnamefont#1{#1}\fi
\expandafter\ifx\csname citenamefont\endcsname\relax
  \def\citenamefont#1{#1}\fi
\expandafter\ifx\csname url\endcsname\relax
  \def\url#1{\texttt{#1}}\fi
\expandafter\ifx\csname urlprefix\endcsname\relax\def\urlprefix{URL }\fi
\providecommand{\bibinfo}[2]{#2}
\providecommand{\eprint}[2][]{\url{#2}}

\bibitem[{\citenamefont{Eelkema et~al.}(2006)\citenamefont{Eelkema, Pollard,
  Vicario, Katsonis, Ramon, Bastiaansen, Broer, and Feringa}}]{Eelkema06}
\bibinfo{author}{\bibfnamefont{R.}~\bibnamefont{Eelkema}},
  \bibinfo{author}{\bibfnamefont{M.~M.} \bibnamefont{Pollard}},
  \bibinfo{author}{\bibfnamefont{J.}~\bibnamefont{Vicario}},
  \bibinfo{author}{\bibfnamefont{N.}~\bibnamefont{Katsonis}},
  \bibinfo{author}{\bibfnamefont{B.~S.} \bibnamefont{Ramon}},
  \bibinfo{author}{\bibfnamefont{C.~W.~M.} \bibnamefont{Bastiaansen}},
  \bibinfo{author}{\bibfnamefont{D.~J.} \bibnamefont{Broer}}, \bibnamefont{and}
  \bibinfo{author}{\bibfnamefont{B.~L.} \bibnamefont{Feringa}},
  \bibinfo{journal}{Nature} \textbf{\bibinfo{volume}{440}},
  \bibinfo{pages}{163} (\bibinfo{year}{2006}).

\bibitem[{\citenamefont{Huang et~al.}(2003)\citenamefont{Huang, Zorman,
  Mehregany, and Roukes}}]{Huang03}
\bibinfo{author}{\bibfnamefont{X.~M.~H.} \bibnamefont{Huang}},
  \bibinfo{author}{\bibfnamefont{C.~A.} \bibnamefont{Zorman}},
  \bibinfo{author}{\bibfnamefont{M.}~\bibnamefont{Mehregany}},
  \bibnamefont{and} \bibinfo{author}{\bibfnamefont{M.~L.}
  \bibnamefont{Roukes}}, \bibinfo{journal}{Nature}
  \textbf{\bibinfo{volume}{421}}, \bibinfo{pages}{496} (\bibinfo{year}{2003}).

\bibitem[{\citenamefont{Okawachi et~al.}(2005)\citenamefont{Okawachi, Bigelow,
  Sharping, Zhu, Schweinsberg, Gauthier, Boyd, and Gaeta}}]{Okawachi05}
\bibinfo{author}{\bibfnamefont{Y.}~\bibnamefont{Okawachi}},
  \bibinfo{author}{\bibfnamefont{M.~S.} \bibnamefont{Bigelow}},
  \bibinfo{author}{\bibfnamefont{J.~E.} \bibnamefont{Sharping}},
  \bibinfo{author}{\bibfnamefont{Z.}~\bibnamefont{Zhu}},
  \bibinfo{author}{\bibfnamefont{A.}~\bibnamefont{Schweinsberg}},
  \bibinfo{author}{\bibfnamefont{D.~J.} \bibnamefont{Gauthier}},
  \bibinfo{author}{\bibfnamefont{R.~W.} \bibnamefont{Boyd}}, \bibnamefont{and}
  \bibinfo{author}{\bibfnamefont{A.~L.} \bibnamefont{Gaeta}},
  \bibinfo{journal}{Phys.~Rev.~Lett.} \textbf{\bibinfo{volume}{94}},
  \bibinfo{pages}{153902} (\bibinfo{year}{2005}).

\bibitem[{\citenamefont{Hartland}(2006)}]{Hartland06}
\bibinfo{note}{For a review, see} \bibinfo{author}{\bibfnamefont{G.~V.} \bibnamefont{Hartland}},
  \bibinfo{journal}{Annu.~Rev.~Phys.~Chem.} \textbf{\bibinfo{volume}{57}},
  \bibinfo{pages}{403} (\bibinfo{year}{2006}).

\bibitem[{\citenamefont{Averitt et~al.}(1997)\citenamefont{Averitt, Sarkar, and
  Halas}}]{Averitt97}
\bibinfo{author}{\bibfnamefont{R.~D.} \bibnamefont{Averitt}},
  \bibinfo{author}{\bibfnamefont{D.}~\bibnamefont{Sarkar}}, \bibnamefont{and}
  \bibinfo{author}{\bibfnamefont{N.~J.} \bibnamefont{Halas}},
  \bibinfo{journal}{Phys.~Rev.~Lett.} \textbf{\bibinfo{volume}{78}},
  \bibinfo{pages}{4217} (\bibinfo{year}{1997}).

\bibitem[{\citenamefont{Oldenburg et~al.}(1998)\citenamefont{Oldenburg,
  Averitt, Westcott, and Halas}}]{Oldenburg98}
\bibinfo{author}{\bibfnamefont{S.~J.} \bibnamefont{Oldenburg}},
  \bibinfo{author}{\bibfnamefont{R.~D.} \bibnamefont{Averitt}},
  \bibinfo{author}{\bibfnamefont{S.~L.} \bibnamefont{Westcott}},
  \bibnamefont{and} \bibinfo{author}{\bibfnamefont{N.~J.} \bibnamefont{Halas}},
  \bibinfo{journal}{Chem.~Phys.~Lett.} \textbf{\bibinfo{volume}{288}},
  \bibinfo{pages}{243} (\bibinfo{year}{1998}).

\bibitem[{\citenamefont{Graf and van Blaaderen}(2002)}]{Graf02}
\bibinfo{author}{\bibfnamefont{C.}~\bibnamefont{Graf}} \bibnamefont{and}
  \bibinfo{author}{\bibfnamefont{A.}~\bibnamefont{van Blaaderen}},
  \bibinfo{journal}{Langmuir} \textbf{\bibinfo{volume}{18}},
  \bibinfo{pages}{524} (\bibinfo{year}{2002}).

\bibitem[{\citenamefont{Graf et~al.}(2003)\citenamefont{Graf, Vossen, Imhof,
  and van Blaaderen}}]{Graf03}
\bibinfo{author}{\bibfnamefont{C.}~\bibnamefont{Graf}},
  \bibinfo{author}{\bibfnamefont{D.~L.~J.} \bibnamefont{Vossen}},
  \bibinfo{author}{\bibfnamefont{A.}~\bibnamefont{Imhof}}, \bibnamefont{and}
  \bibinfo{author}{\bibfnamefont{A.}~\bibnamefont{van Blaaderen}},
  \bibinfo{journal}{Langmuir} \textbf{\bibinfo{volume}{19}},
  \bibinfo{pages}{6693} (\bibinfo{year}{2003}).

\bibitem[{\citenamefont{Portales et~al.}(2002)\citenamefont{Portales, Saviot,
  Duval, Gaudry, Cottancin, Pellarin, Lerm\'{e}, and Broyer}}]{Portales02}
\bibinfo{author}{\bibfnamefont{H.}~\bibnamefont{Portales}},
  \bibinfo{author}{\bibfnamefont{L.}~\bibnamefont{Saviot}},
  \bibinfo{author}{\bibfnamefont{E.}~\bibnamefont{Duval}},
  \bibinfo{author}{\bibfnamefont{M.}~\bibnamefont{Gaudry}},
  \bibinfo{author}{\bibfnamefont{E.}~\bibnamefont{Cottancin}},
  \bibinfo{author}{\bibfnamefont{M.}~\bibnamefont{Pellarin}},
  \bibinfo{author}{\bibfnamefont{J.}~\bibnamefont{Lerm\'{e}}},
  \bibnamefont{and} \bibinfo{author}{\bibfnamefont{M.}~\bibnamefont{Broyer}},
  \bibinfo{journal}{Phys.~Rev.~B} \textbf{\bibinfo{volume}{65}},
  \bibinfo{pages}{165422} (\bibinfo{year}{2002}), \eprint{cond-mat/0203468}.

\bibitem[{\citenamefont{Mazurenko et~al.}(2004)\citenamefont{Mazurenko, Moroz,
  Graf, van Blaaderen, and Dijkhuis}}]{MySPIEAu}
\bibinfo{author}{\bibfnamefont{D.~A.} \bibnamefont{Mazurenko}},
  \bibinfo{author}{\bibfnamefont{A.}~\bibnamefont{Moroz}},
  \bibinfo{author}{\bibfnamefont{C.~M.} \bibnamefont{Graf}},
  \bibinfo{author}{\bibfnamefont{A.}~\bibnamefont{van Blaaderen}},
  \bibnamefont{and} \bibinfo{author}{\bibfnamefont{J.~I.}
  \bibnamefont{Dijkhuis}}, \bibinfo{journal}{Proc. SPIE Int. Soc. Opt. Eng.}
  \textbf{\bibinfo{volume}{5450}}, \bibinfo{pages}{569} (\bibinfo{year}{2004}).

\bibitem[{\citenamefont{Kelf et~al.}(2005)\citenamefont{Kelf, Sugawara,
  Baumberg, Abdelsalam, and Bartlett}}]{Kelf05}
\bibinfo{author}{\bibfnamefont{T.~A.} \bibnamefont{Kelf}},
  \bibinfo{author}{\bibfnamefont{Y.}~\bibnamefont{Sugawara}},
  \bibinfo{author}{\bibfnamefont{J.~J.} \bibnamefont{Baumberg}},
  \bibinfo{author}{\bibfnamefont{M.}~\bibnamefont{Abdelsalam}},
  \bibnamefont{and} \bibinfo{author}{\bibfnamefont{P.~N.}
  \bibnamefont{Bartlett}}, \bibinfo{journal}{Phys.~Rev.~Lett.}
  \textbf{\bibinfo{volume}{95}}, \bibinfo{pages}{116802}
  (\bibinfo{year}{2005}), \eprint{physics/0505061}.

\bibitem[{\citenamefont{Mazurenko}(2004)}]{Mythesis}
\bibinfo{author}{\bibfnamefont{D.~A.} \bibnamefont{Mazurenko}}, \href{http://igitur-archive.library.uu.nl/dissertations/2004-1025-1%
25904/index.htm}{\underline{Ph.D. thesis}},
  \bibinfo{school}{Universiteit Utrecht} (\bibinfo{year}{2004}).

\bibitem[{\citenamefont{Plech et~al.}(2004)\citenamefont{Plech, Kotaidis,
  Gr\'{e}sillon, Dahmen, and {von Plessen}}}]{Plech04}
\bibinfo{author}{\bibfnamefont{A.}~\bibnamefont{Plech}},
  \bibinfo{author}{\bibfnamefont{V.}~\bibnamefont{Kotaidis}},
  \bibinfo{author}{\bibfnamefont{S.}~\bibnamefont{Gr\'{e}sillon}},
  \bibinfo{author}{\bibfnamefont{C.}~\bibnamefont{Dahmen}}, \bibnamefont{and}
  \bibinfo{author}{\bibfnamefont{G.}~\bibnamefont{{von Plessen}}},
  \bibinfo{journal}{Phys.~Rev.~B} \textbf{\bibinfo{volume}{70}},
  \bibinfo{pages}{195423} (\bibinfo{year}{2004}).

\bibitem[{\citenamefont{Lamb}(1882{\natexlab{a}})}]{Lamb82sh}
\bibinfo{author}{\bibfnamefont{H.}~\bibnamefont{Lamb}},
  \bibinfo{journal}{Proc.~Lond.~Math.~Soc.} \textbf{\bibinfo{volume}{14}},
  \bibinfo{pages}{50} (\bibinfo{year}{1882}{\natexlab{a}}).

\bibitem[{\citenamefont{Duval}(1992)}]{Duval92}
\bibinfo{author}{\bibfnamefont{E.}~\bibnamefont{Duval}},
  \bibinfo{journal}{Phys.~Rev.~B} \textbf{\bibinfo{volume}{46}},
  \bibinfo{pages}{5795} (\bibinfo{year}{1992}).

\bibitem[{\citenamefont{Lamb}(1882{\natexlab{b}})}]{Lamb82sp}
\bibinfo{author}{\bibfnamefont{H.}~\bibnamefont{Lamb}},
  \bibinfo{journal}{Proc.~Lond.~Math.~Soc.} \textbf{\bibinfo{volume}{13}},
  \bibinfo{pages}{189} (\bibinfo{year}{1882}{\natexlab{b}}).

\bibitem[{\citenamefont{Courty et~al.}(2002)\citenamefont{Courty, Lisiecki, and
  Pileni}}]{Courty02}
\bibinfo{author}{\bibfnamefont{A.}~\bibnamefont{Courty}},
  \bibinfo{author}{\bibfnamefont{I.}~\bibnamefont{Lisiecki}}, \bibnamefont{and}
  \bibinfo{author}{\bibfnamefont{M.~P.} \bibnamefont{Pileni}},
  \bibinfo{journal}{J.~Chem.~Phys.} \textbf{\bibinfo{volume}{116}},
  \bibinfo{pages}{8074} (\bibinfo{year}{2002}).

\bibitem[{\citenamefont{Bachelier and Mlayah}(2004)}]{Bachelier04}
\bibinfo{author}{\bibfnamefont{G.}~\bibnamefont{Bachelier}} \bibnamefont{and}
  \bibinfo{author}{\bibfnamefont{A.}~\bibnamefont{Mlayah}},
  \bibinfo{journal}{Phys.~Rev.~B} \textbf{\bibinfo{volume}{69}},
  \bibinfo{pages}{205408} (\bibinfo{year}{2004}).

\bibitem[{\citenamefont{Maznev et~al.}(1997)\citenamefont{Maznev, Hohlfeld, and
  G\"{u}dde}}]{Maznev97}
\bibinfo{author}{\bibfnamefont{A.~A.} \bibnamefont{Maznev}},
  \bibinfo{author}{\bibfnamefont{J.}~\bibnamefont{Hohlfeld}}, \bibnamefont{and}
  \bibinfo{author}{\bibfnamefont{J.}~\bibnamefont{G\"{u}dde}},
  \bibinfo{journal}{J.~Appl.~Phys.} \textbf{\bibinfo{volume}{82}},
  \bibinfo{pages}{5082} (\bibinfo{year}{1997}).

\bibitem[{\citenamefont{Garfinkel et~al.}(1966)\citenamefont{Garfinkel,
  Tiemann, and Engeler}}]{Garfinkel66}
\bibinfo{author}{\bibfnamefont{M.}~\bibnamefont{Garfinkel}},
  \bibinfo{author}{\bibfnamefont{J.~J.} \bibnamefont{Tiemann}},
  \bibnamefont{and} \bibinfo{author}{\bibfnamefont{W.~E.}
  \bibnamefont{Engeler}}, \bibinfo{journal}{Phys.~Rev.}
  \textbf{\bibinfo{volume}{148}}, \bibinfo{pages}{695} (\bibinfo{year}{1966}).

\bibitem[{\citenamefont{Nix and MacNair}(1941)}]{Nix41}
\bibinfo{author}{\bibfnamefont{F.~C.} \bibnamefont{Nix}} \bibnamefont{and}
  \bibinfo{author}{\bibfnamefont{D.}~\bibnamefont{MacNair}},
  \bibinfo{journal}{Phys.~Rev.} \textbf{\bibinfo{volume}{60}},
  \bibinfo{pages}{597} (\bibinfo{year}{1941}).

\bibitem[{\citenamefont{Hartland}(2002)}]{Hartland02}
\bibinfo{author}{\bibfnamefont{G.~V.} \bibnamefont{Hartland}},
  \bibinfo{journal}{J.~Chem.~Phys.} \textbf{\bibinfo{volume}{106}},
  \bibinfo{pages}{8048} (\bibinfo{year}{2002}).

\end{thebibliography}

\end{document}